\documentclass[prl,twocolumn,amsmath,amssymb,floatfix,showpacs,superscriptaddress]{revtex4}

\usepackage{bm}

\usepackage{amssymb,amsmath}
\usepackage{graphics}
\usepackage{epstopdf}
\usepackage{epsfig}
\usepackage{color}
\usepackage{amsfonts}
\usepackage{bm}
\usepackage{amscd}
\usepackage{epsfig}
\usepackage{subfigure}
\usepackage{color}

\newcommand{\be}{\begin{equation}}
\newcommand{\ee}{\end{equation}}

\begin{document}
\title{Superconducting fluctuations--Large Diamagnetism of Low $T_c$ Nanoparticles}

\author{ Yoseph  Imry}
\affiliation{Department of Condensed Matter Physics, Weizmann Institute of Science, Rehovot 76100, Israel}.

\date{\today}


\begin{abstract}

It is shown that nanoparticles made of low $T_c$ superconductors have large diamagnetic response
at temperatures  several orders of magnitude above $T_c$.
Most features of the recently observed Giant diamagnetism of Au nanorods are explained in terms of superconducting fluctuations, except for the huge magnitude of the effect.

\end{abstract}

\maketitle

Very recently \cite{Giant-13}, a novel nanoscale effect-- a large average diamagnetic susceptibility of rod-shaped, down to ten-nanometer scale, gold nanoparticles-- has been discovered experimentally. Such a susceptibility should be due to persistent currents \cite{book} flowing in these nanoparticles in response to a magnetic field. In fact,  Ref. \cite{Giant-14} proposed an explanation of this effect in terms of the persistent currents flowing on the surfaces of these nanoparticles in response to the magnetic flux,  using a model of ballistic, noninteracting electrons. According to Refs \cite{Giant-13,Giant-14}, the effect of  Ref.\cite{Giant-13} is intrinsic to the metal, and not due to chemical interactions with  a capping layer.

The experience \cite{HG,book,av} in explaining such mesoscopic currents shows however, that just finite-size effects due to  noninteracting electrons fall short in explaining them both in sign and in magnitude. The reason being the alternating sign of the response as function of the azimuthal quantum number. This yields a persistent  current whose sign varies from sample to sample (due to disorder, and/or to minute changes in, say,  the sample's radius). The resulting average over an ensemble of many samples becomes very small. In fact, this average is on the order of the level spacing \cite{av}, while the required persistent current \cite{Giant-14} is of the order of the Thouless energy. The ratio of the latter to the former is on the order of several hundreds for a compact nano-particle 
with a linear size 10nm and a comparable mean free path \cite{mfp}. Therefore, electron-electron interactions must be invoked  to give the current a definite sign and to account for the average current
\cite{AL-74,AE0,AE,book}. The diamagnetic sign of the response demands attractive interactions, as in a superconductor. 

This work is motivated by the above experimental results, but we believe that this study leads to a much more general insight: as may be expected on general grounds, the effect of fluctuations increases with decreasing sample size. On the nanoscale, especially in superconductors with their large coherence lengths,  fluctuations may become dominant over the averages!.

The model we use here, invokes superconducting fluctuations, {\em much above $T_c$}, of the conduction electrons. We state from the outset that, for gold, it gives "only" about an order of magnitude increase of the susceptibility compared to $\chi_L$ (the Landau diamagnetic susceptibility of the conduction electrons). The results of Ref. \cite{Giant-13} are {\em three orders of magnitude above $\chi_L$}. Thus, they are just an example and provide motivation.
Although the strength of superconducting fluctuations at such high temperatures is a truly  general and remarkable phenomenon, and it otherwise 
explains all other features of the data, including the very weak temperature dependence up to $\sim 10^5 T_c$, something (as far as explaining the results of Ref \cite{Giant-13}) is still missing! 

 The inherent difficulty of this problem stems from the fact that the bulk volume susceptibility of Au, $\chi_b \sim$ several $\sim10^{-5}$,  results from the dense core electrons, which {\it should not change very much  with the arrangement and binding of the atoms (for example, in the metal or the nanoparticle).} The electrons that do  change and therefore should be expected to yield the effect, are the valence/conduction ones, whose Landau susceptibility, $\chi_L$, is $\sim$  2 orders of mag smaller!
Thus, to explain the observed nanorod effect, a susceptiblity larger by roughly an order of magnitude than $\chi_b$, one needs, as stated,  a {\it $\sim 3$ orders of magnitude
boost over $\chi_L$}.

Here we start with the finding of Ref. \cite{we} that the magnitude of the persistent currents in Au (and other noble metals)  is explainable
assuming that when they are pure bulk they are {\em superconductors with $T_c$ on the scale of a  $mK$ or a fraction thereof \cite{pairbr}}. We shall see that  the  same  assumption about the superconductivity of these metals, when pure, qualitatively explains all trends of the giant diamagnetic susceptibility of Au nanorods as well. The mechanism being superconducting fluctuations much above $T_c$.  However, as  stated, this explanation still falls short by about two orders of magnitude (out of three) in yielding the magnitude of the giant diamagnetism. We shall also mention here the further change of sign of the susceptibility for the even smaller size range. \cite{we-13,REICH,para}. This actually is in agreement with the theoretical picture \cite{we-13}.

A set of 10 colloidal spherically-capped Au nanorod systems was prepared in Ref.\cite{Giant-13}. They were single-crystalline, with an electronic mean-free path similar to the bulk ($\sim$ 60nm). Their radii ranged from 7 to 31 nm, and aspect ratios from 2.4 to 7. Due to the large anisotropy of the magnetic susceptibility, the rods were aligned by a large (33 T) magnetic field with cylinder axis parallel to the magnetic field. The alignment was confirmed by the anisotropic optical response to polarized light.
Magnetic-field induced linear dichroism and birefringence were induced by the field and yielded  the magnetic susceptibilities 
parallel ($\chi_\parallel $) and perpendicular ($\chi_\perp $)  to the cylinder axis and their difference, $\Delta \chi_ V >0$. These were confirmed by SQUID measurements.

The susceptibilities were negative (diamagnetic), increasing with decreasing size, larger than that of the bulk by an order of magnitude (depending on size and aspect ratio) and temperature-independent in the whole measurement range of 5-300K.
We emphasize that $|\chi_\perp |$ is larger than $|\chi_{\parallel}|$. Their ratio increases with the aspect ratio of the cylinder.

These results are rather unexpected and quite difficult to understand, especially the huge size and the temperature independence of a mesoscopic effect. Here we show that the superconducting fluctuations much above $T_c$ qualitatively explain, except for the already mentioned order of magnitude, {\em all} the trends  of these results.
More generally, the importance of fluctuations on the nanoscale is highlighted.

The most straightforward way to understand these effects qualitatively is by employing the Physical picture of Schmid \cite{AS-69}, based on the Ginzburg-Landau (G-L) theory for the fluctuations. His results for bulk 2D and 3D systems are consistent with those of the microscopic  calculations \cite{HS-68,AL-74}. We believe that this theory, with appropriate parameters (dropping the approximation of being close to $T_c$) is qualitatively valid even much above $T_c$. There, actually, the gaussian approximation (retaining only quadratic terms in the order parameter) is very well valid. Moreover, as long as the dimensions of the nanoparticle are much smaller than the relevant coherence length (see below), only fluctuations in which the order parameter, $\psi$ is uniform  over the whole nanoparticle, matter \cite{0D,Doug}. The gaussian  free energy density of a such a  fluctuation is given by $a |\psi|^2$, where $a$ is the appropriate G-L parameter. The evaluation of the integral over the 0D fluctuations was done in Refs \cite{0D,Doug}, and far above $T_c$  it reduces to the gaussian approximation. So does the full calculation of the susceptibility \cite{Doug}. We adopt, following Schmid, the normalization of $\psi$ where $|\psi|^2$ is  the fluctuating superfluid density. Then, $a = \hbar ^2 /(2m \xi(T) ^2)$, where $\xi(T)$ is the coherence length in the bulk. For the nanorod volume $\cong \pi R^2 L$, the total free energy of the flucutuation is $\cong \pi R^2 La |\psi|^2$. This (over the temperature $T$) sets the {\em gaussian} probability for the fluctuation,  which implies that the average fluctuating superfluid density is
\be
<|\psi|^2>  = \frac{k_BT}{2 \pi R^2 La}  =  \frac{k_BT m \xi(T) ^2}{ \pi R^2 L \hbar ^2}.
\ee 
We shall later use
$k_{\rm B}=1$.

Adopting the Langevin expression for the diamagnetic susceptibility per unit volume of a finite, mobile-charge carrying, entity
\be
\chi_{d,L} = \frac{nq^2 <r^2>}{4mc^2},
\ee
where $n$ is the density of charge carriers, $q$ their charge,   $m$ their mass and $<r^2>$ their typical radius-of-motion squared.
For $\xi(T)$ we take the "normal-metal coherence length" which agrees for $T  >>T _c$ with the GL length . It is also the characteristic scale for interaction effects \cite{book}. For a dirty metal. \cite{PdG}
\be
\xi^2(T) = \frac  {\pi \hbar D}{8T}.
\ee
For $T \sim T_c$ this yields the dirty limit $T=0$ G-L coherence length, which is of the order of 1000nm for the gold used in Ref.\cite{Giant-13}.
Putting the above together we get for $\chi_d$ much above $T_c$
\be
\chi_d = \frac{ e^2 D <r^2>}{ 8 \hbar c^2 R^2 L  }. \label{AS}
\ee
Expressing this  in terms of the Landau susceptibility for a normal metal, $\chi_L = \frac{e^2 k_F}{12 \pi^2 mc^2}$ we find,
allowing for $D = v_F \ell / 3$, where $\ell$ is the elastic mean free path,
\be
\frac{\chi_d}{\chi_L}= \frac{\pi^2 \ell<r^2>}{ 2R^2 L  }.
\ee
Taking the typical orbit radii $<r^2>_{parallel,perp}$ to be $A_{parallel}R^2$ and $A_{perp}RL$ where $A_{parallel,perp}$ are numerical constants of order unity, and the indices parallel and  perp referring to  the directions of the magnetic field vs. the cylinder's axis, we find
\be
\frac{\chi_{d, parallel}}{\chi_L} = \pi^2 A_{parallel} \ell/2L; \frac{\chi_{d, perp}}{\chi_L} = \pi^2 A_{perp} \ell /2R.
\ee
In the the bulk $\ell \cong 60 nm$ and because the rods  being single-crystalline \cite{Giant-13}, they should have values of $\ell$, similar to that of the bulk. Thus,  both $\chi_d$'s are larger 
than $\chi_L$ by sizable numerical factors which {\it increase} with decreasing nanocylinder size. Moreover $\chi_{d, perp}$ is larger than $\chi_{d, parallel}$ by the aspect ratio, $L/R$,
of the cylinder. 
The most remarkable feature of these simple results is the temperature independence, which simply follows from the {\em cancellation of the T factor of the fluctuations (equipartition theorem) and the $1/T$ one of $\xi^2(T)$}. As we shall see, this is valid for temperatures below the effective Thouless energy.

The two main features which appear in the microscopic theory and are neglected in the simplest classical (static) and uniform ($q=0$) fluctuation theory are the finite wavenumber, $q$ and Matsubara frequency ($\omega_{\nu} = \nu (2 \pi T)$, with $\nu$ being an integer). As to the former, we note \cite{0D,Doug} that the reason that finite wavenumber, $q$, fluctuations are expected to be negligible (at low temperature) for our nano-system, is the following: in a dirty superconductor, $D/ \xi^2 \sim T_c$. For Au, our estimate for $T_c$ is a fraction of a mK  (and much smaller estimates exist) and the L's of Ref. \cite{Giant-13} are on the order of 10nm, which leads to $D/ L^2 \cong 100-200K$ (the Thouless energy). Thus, at $10T_c$, the smallest energy of a nonzero q fluctuation is larger than $T$ by four orders of magnitude! As to the latter, these quantum fluctuations are not expected to be important at temperatures much above $T_c$. They may still produce significant corrections for very small systems \cite{Levinson}, as we shall see below. 

To understand the essence of the differences between the fluctuation G-L  and the microscopic  theory, we compare the results for the paradigmatic case of the orbital response of a thin small \cite{book,AE} ring to a magnetic field, or flux.
For a thin ring of radius R and small height L, adaptation of the Schmid \cite{AS-69} approach, as in Eq. \ref{AS} gives:
\be
\chi_{d,GL} = \frac{ e^2 D }{ 8 \hbar c^2  L  }. \label{AS1}
\ee
Ref \cite{AE} calculated the persistent current of such a ring, using the microscopic perturbation theory. We get the magnetic moment by multiplying with $\pi R^2/c$ and hence, for the dominant first harmonic in the flux
\be
\chi_{d,AE}= \frac{ 4 e^2 D }{ \pi^2 \hbar c^2  L ln(T_1/T_c)}, \label{AE}
\ee
where $T_1= \frac{\hbar D}{(2 \pi R)^2}$ is the Thouless energy.  Both results are for $T \lesssim T_1$.
We see that the microscopic result is approximately given by the fluctuation G-L one multiplied by  $32/ (\pi^2ln(T_1/T_c)) \sim 1/4$ for the Au samples of Ref.\cite{Giant-13}. The $1/ln$ factor describes (see below) the renormalized  attractive (below the Debye energy) interaction at the "Physical scale" $T_1$. This is the relevant scale for these mesoscopic phenomena \cite{book,we} at  temperaures $\lesssim T_1$. With this reduction, the susceptibility, especially the perpendicular one, can still be larger, but now by more modest factors, than $\chi_L$. All trends of the G-L results are  of course still satisfied. This includes the unusual temperature independence below $T_1$.

A very satisfying  feature of the microscopic result of Eq \ref{AE} is the, albeit weak, dependence on $T_c$. $\chi$ vanishing as
$\frac{1}{ln(T_1/T_c)}$ with $T_c$. $T_c = 0$ is the normal metal limit.

To explain the scale-dependence of the interaction, we recall briefly how it is derived.
By integrating over thin shells in momentum (or energy) space, one obtains  the well-known (see e.g. \cite{MA,PdG,we-14}) variation of the electron-electron interaction coupling $g$, be it repulsive or attractive, from 
a high-energy scale $\omega_{>}$ 
to a low-energy one $\omega_{<}$,
\begin{equation}
\frac{1}{g(\omega^{}_{<})}=\frac{1}{g(\omega^{}_{>})}+\log\Bigl(\frac{\omega^{}_{>}}{\omega^{}_{<}}\Bigr)\ .\label{RG}
\end{equation}
Notice that a repulsive/attractive interaction is ``renormalized downwards/upwards"  with decreasing energy scale $\omega^{}_<$.
What makes superconductivity possible is that at $\omega_{\rm D}$ the renormalized repulsion is much  smaller than its value on the microscopic scale. 
At $\omega_{\rm D}$ the attraction may  win and then at lower energies the total interaction increases in absolute value, until it diverges at some small energy scale, 
the conventional  $T^{}_c$ of the given material. Choosing $ \omega^{}_{<} = T_c$ (where the inverse interaction vanishes) and $\omega^{}_{>}$ to be the "Physical scale", this gives
\begin{equation}
\frac{1}{g(\omega^{}_{>})} = ln (\frac{\omega^{}_{>}}{T_c})
\end{equation}
The physical scale for the (dominant) first moment of the flux-dependence of the persistent current in a ring is the Thouless energy, $T_1$ (in the notation of  Ref.\cite{AE}). Thus, the $1/ln(\frac{T_1}{T_c})$ factor in the AE result is just the appropriate renormalized interaction, replacing the bare interaction of Ref.\cite{AE0}, as hinted in Ref. \cite{AE}. This interaction is attractive for a superconductor when the Physical scale is below the Debye energy. However, it should change sign for physical scales above $\cong \omega_D $ \cite{we-13}.

For $T_1 << T$, the physical scale becomes T. This gives the usual G-L temperature dependence of the various quantities, especially relevant when $T $ approaches $T_c$.

We mention that the above change of sign has serious consequences for the magnetic response.
As mentioned in Ref. \cite{Giant-13}, in the even smaller size-range (a few nm), gold nanoparticles become paramagnetic \cite{REICH,para}. This is not treated here. However, it should be mentioned that  this change of sign was explained in Ref \cite{we-13} in terms of the scale-dependence of the renormalized interaction, as briefly mentioned above. Very interestingly, then, when noble (and other, low $T_c$) metals nanoparticles  decrease in size towards  the 10 nm scale, their average diamagnetic susceptibility becomes stronger. Further decrease in size, to the few nm scale, will give a change to a paramagnetic orbital response.
All this is very qualitatively consistent with existing experiments. Systematic examination of this rich behavior, for nanoparticles of the same material as function of size, should be instructive.

We conclude this note with a speculation on the origin of the giant diamagnetic susceptibility \cite{Giant-13}. Its order of magnitude is on a scale that suggests the importance of the dense atomic cores. Can these be sensitive to superconducting correlations of the conduction electrons? This might be due to a proximity effect between these two types of electrons.

\begin{acknowledgements}

I thank Yuval Oreg, Alexander Finkelstein, Ora Entin-Wohlman, Amnon Aharony and  Erez Berg for important discussions, Hamutal Bary-Soroker for participation in and contributions to Refs 
\cite{we,we-13} and P.C.M. Chiristianen for helpful correspondence.
This work was supported by the Israeli Science Foundation (ISF) and the US-Israel Binational Science Foundation (BSF). The author first learned of this problem from an instructive   lecture by A. Hernando in Coma-Ruga 14, 10th International Workshop on Nanomagnetism and Superconductivity at the Nanoscale.

\end{acknowledgements}


\begin{thebibliography}{999}



\bibitem{Giant-13}
P. G. van Rhee.,  P. Zijlstra, T.G.A Verhagen,J. Aarts, M.J. Katsnelson, J.C. Maan, M. Orrit and P.C.M. Christianen, Phys Rev Lett. {\bf 111}, 127202 (2013).

\bibitem{book}
Y. Imry, {\it Introduction to Mesoscopic Physics}, 2nd ed.
(Oxford University Press, Oxford, 2002).



\bibitem{Giant-14} A. Hernando, A Ayuela,
P Crespo and P M Echenique, New J. Phys. {\bf 16},  073043 (2014).


\bibitem{HG} H. Bouchiat and G. Montambaux, J. Phys. (Paris) {\bf 50}, 2695 (1989).

\bibitem{av} B.L. Altshuler, Y. Gefen and Y. Imry. 
Phys. Rev. Lett. {\bf 66}, 88 (1991).

\bibitem{mfp} Even if the bulk had a much larger mean free path, in a realistic nanoparticle any surface imperfection should limit the mean free path to the nanoparticle's typical size. This was first understood in a related problem by:  K. Fuchs. The conductivity of thin metallic films according to
the electron theory of metals, Proc. Camb. Phil. Soc, {\bf 34}, 100 (1938).




\bibitem{AL-74} L. G. Aslamazov and A. I. Larkin, Sov. Phys. JETP, {\bf 40}, 321 (1975).

\bibitem{AE} V. Ambegaokar and U. Eckern, Europhys. Lett {\bf 13}, 733 (1990).

\bibitem{AE0} V. Ambegaokar, U. Eckern,
Phys. Rev. Lett. 65, 381-384 (1990)



\bibitem{we}
H. Bary-Soroker, O. Entin-Wohlman, and Y. Imry, Phys. Rev. Lett.
{\bf 101}, 057001 (2008);
H. Bary-Soroker, O. Entin-Wohlman, and Y. Imry, 
Phys. Rev. B {\bf
80}, 024509 (2009) and references therein.

\bibitem{AS-69} A. Schmid, Phys Rev {\bf 180}, 527 (1969).

\bibitem{HS-68} H. Schmidt, Z. Physik {\bf 216}, 336 (1968).

\bibitem{0D} V. V. Shmidt, JETP Lett {\bf 3}, 89 (1966) and in: proceedings of the Tenth International Conference on Iow-Temperature Physics, Moscow, 1966, edited by M. P. Malkov , Vol. 2B, p. 205 (1967). 

\bibitem{Doug}  B. M\"{u}hlschlegel, D J Scalapino and R Denton, Phys Rev {\bf B}6, 1767 (1972).

\bibitem{we-13} H. Bary-Soroker, O. Entin-Wohlman, Y. Imry and A. Aharony, Phys Rev Lett {\bf 110}, 056801 (2013),

\bibitem{REICH}
S. Reich, G. Leitus, and Y. Feldman, , Appl. Phys. Lett. {\bf 88}, 222502 (2006) and references therein. 

\bibitem{para} Y. Yamamoto, T. Miura, M. Suzuki, N. Kawamura, H.Miyagawa, T. Nakamura, K. Kobayashi, T. Teranishi, and H. Hori, ,  Phys. Rev. Lett. {\bf 93}, 116801 (2004);
Y. Negishi, H. Tsunoyama, M. Suzuki, N. Kawamura, M.M. Matsushita, K. Maruyama, T. Sugawara, T. Yokoyama, and T. Tsukuda, , J. Am. Chem. Soc. {\bf 128}, 12 034 (2006);
J. Bartolome, F. Bartolome, L. M. Garcia, A. I. Figueroa, A. Repolle, M. J. Martinez, F. Luis, C. Magen, S. Selenska-Pobell, F. Pobell, T. Reitz, R. Schšnemann, T. Herrmannsdšrfer, M. Merroun, A. Geissler, F. Wilhelm, and A. Rogalev,  Phys. Rev. Lett. {\bf 109}, 247203 (2012);
 P. Crespo, R. Litra ´n, T. C. Rojas, M. Multigner, J. M. de la Fuente, J. C. Sanchez-Lopez, M. A. Garcia, A. Hernando, S. Penades, and A. Fernandez,  Phys. Rev. Lett. {\bf 93}, 087204 (2004);
 J. S. Garitaonandia, M. Insausti, E. Goikolea, M. Suzuki, J. D. Cashion, N. Kawamura, H. Ohsawa, I. Gil de Muro, K. Suzuki, F. Plazaola {\it et al.}, , Nano Lett. {\bf 8}, 661 (2008);
 R. Gr\'{e}get
et al. ChemPhysChem {\bf 13}, 3092 (2012).
 
\bibitem{pairbr} According to Ref \cite{we}, the reason that superconductivity has never directly been  observed in the noble metals
is a minute (sub-ppm) amount of pair-breakers which are very difficult to avoid in real samples, but their concentration, although enough to reduce the $T_c$ of real samples to below the measurable range, is  too small to affect the mesoscopic persistent currents  at the relevant range.






\bibitem{we-14}
A. Aharony, O. Entin-Wohlman, and Y. Imry (2014), to be published in J. Stat. Phys., a Memorial Volume for K. G. Wilson.





\bibitem{PdG}  P.G. de Gennes,  Superconductivity of Metals and Alloys, Benjamin (1964).

\bibitem{Levinson} A. Aharony, O. Entin-Wohlman, H. Bary-Soroker, and Y. Imry, Lithuanian Journal of Physics {\bf 52}, 81 (2012).

\bibitem{MA}
P. Morel and P. W. Anderson, Phys. Rev. {\bf 125}, 1263 (1962);
N. N. Bogoliubov, V. V. Tolmachev, and D. V. Shirkov, {\it A New
Method in the Theory of Superconductivity} (Consultants Bureau,
Inc., New York, 1959).





\end{thebibliography}
\end{document}